# Dealing with Big Data

Tobias Blanke and Andrew Prescott

## BIG DATA AND ITS ANXIETIES

As the United States grew rapidly in the late nineteenth century, its government sought to collect more information about the state of the country. The vast amount of data collected in the 1880 census was still being transcribed by clerical staff nine years later. A new method was required for the 1890 census and Herman Hollerith, a census employee, proposed that census returns should be analysed by the use of punched cards sorted by an electro-mechanical machine. Operators made holes in the cards which corresponded to (for example) the number of people in a family, and the cards were automatically sorted to provide statistics on family size. Using Hollerith's machines, between 12.5 and 15 million individual records were processed in less than two years in the 'first truly mass-scale state information processing project to be mechanised' (Driscoll 2012: 8). Hollerith's 'Tabulating Machine Company' went on to form part of IBM, and by the 1920s automated information processing was used for US government processes ranging from income tax to fingerprinting of criminals, with other governments following the US lead (Driscoll 2012; Agar 2003).

The potential and perils of automated data processing have fascinated governments and the governed ever since. During the First World War, the British government discussed linking all government registers into a vast national register but was deterred by concerns about civil liberty (Agar 2003). Other public bodies, such as libraries, adopted punched cards and made them familiar to a wider public. Given the long history of automated data processing, it is surprising in 2015 to find government ministers such as Francis Maude, the British Cabinet Office Minister, describing data as 'the new raw material of the 21st century'.[1] Big data has caused similar excitement among academic researchers, with Marshall declaring that 'Big Data is surely the Gold Rush of the Information Age' (Marshall 2012: 1). This excitement reflects two recent developments: the growing quantities of data in machine-readable form which can be processed without extensive preparation; and the availability of advanced methods to analyse this data.

Two examples quoted by Marshall illustrate why big data generates such enthusiasm. The first is Twitter (see also Chapter 5 in this volume), where each tweet is short but has a great deal of metadata associated with it describing the time, place, language, author, etc., of the message.[2] Twitter archives are highly suitable for data mining and can be readily used to create maps showing linguistic groups in a city[3] or visualisations of Russian political groupings (Kell et al. 2012). While such Twitter mining offers exciting research possibilities, the Twitter archive only extends back to its launch in 2006, limiting its relevance for many humanities scholars, while its evidential value is increasingly compromised by the purchase by political and other figures of non-existent followers and by the automated generation of tweets.[4] Marshall's second illustration of the research potential of big data techniques is the Google Ngram Viewer,[5] which shows word trends from millions of books digitised by Google. The Ngram Viewer enables the public for the first time to run their own linguistic analysis of the millions of words in printed books. The inventors of Google NGram Viewer declared that: 'The Ngram Viewer's consequences will transform how we look at ourselves . . . Big data is going to change the humanities, transform the social sciences, and renegotiate the relationship between the world of commerce and the ivory tower' (Aiden and Michel 2013).

Despite the enthusiasm of its creators, there is little evidence at present that the Google Ngram Viewer is generating major new research findings. Gibbs and Cohen (2011) used Google Ngram to analyse references to religion to investigate the crisis of faith and growth of secularisation in the Victorian period. They found that from 1851 there was a decline in the use of the words 'God' and 'Christian' in the titles of books, suggesting that perhaps the Victorian crisis of faith began earlier than previously thought. But Gibbs and Cohen admit that their experiments were very preliminary and Underwood has pointed out that a different selection of words can give a completely opposite result.[6] The hazards of Google Ngram Viewer are further illustrated by a study of the use of the words 'courtesy' and 'politeness' in the eighteenth and nineteenth centuries. A Google Ngram search seemed to show that there was a huge growth in the use of these two words in about 1800. However, until 1800 the letter 's' was frequently printed in a long form so that it looked like an 'f', meaning that these words looked like 'courtefy' and 'politeneff', so the Ngram Viewer missed them. This practice ceased around 1800, so the apparent rise in use of these words reflects typographical changes, not cultural shifts (Jucker 2012: addendum). Other humanities researchers have complained that Ngram results are trivial (Kirsch 2014), and it is certainly true that Ngram often simply provides a visual representation of what we already know. It is not yet clear how we can interrogate Ngram in such a way as to generate new intellectual insights and ideas. It is probable that, in using the big data created by Google Books, researchers will come to prefer more sophisticated search tools, such as that developed by Mark Davies at Brigham Young University.[7]

Big data has many definitions (Jacobs 2009) which often only agree that big data cannot be defined. Some link big data to data sets of a certain size in the petabyte and exabyte range.[8] However, there is no objective reason for categorising these byte ranges as 'big'. Handling a terabyte was problematic ten years ago but is now routine. Big data cannot be seen simply as data that is bigger than the scales we currently experience (Jacobs 2009). For big data, size matters, but the experience of size can only ever be relative. The format of the data is also an issue; large quantities of alphanumeric data may be easier to process than multimedia data. In 'Pathologies of Big Data' (Jacobs 2009) a thought experiment is used to explore the storage of simple demographics about every person on the planet. In the end, there will be 'a table of 6.75 billion rows and maybe 10 columns'. While for humans, this might be an impressive number of data items, such quantities are not a problem for standard computers to process. On the other hand, 8.5 million images of medieval manuscripts[9] can be more difficult for computers to process than eight billion genome sets (for reasons we will explore) and it may be that humans can deal more readily with such big image data. It is impossible to define big data in terms of a particular number of bytes, because 'as technology advances over time, the size of data sets that qualify as big data will also increase' (Chui et al. 2011: 1). We need to think about big data in terms of what computers can store and process and what humans can analyse using current techniques. Big data 'should be defined at any point in time as data whose size forces us to look beyond the tried-and true methods that are prevalent at that time' (Jacobs 2009). Data is big not just in terms of its quantity, but also in terms of what we would like to do with it, how we seek to extract research value from it and the type of infrastructures available to support this extraction of value. As Lynch (2008) emphasises, data can also be big in different ways, making demands in such areas as long-term preservation or description which challenge current computational methods.

Doug Laney's celebrated 2001 'three Vs' description of big data provides a good framework for further analysis: 'Big data is high volume, high velocity, and/or high variety information assets that require new forms of processing to enable enhanced decision making, insight discovery and process optimization' (Laney 2001: 1). High Volume, or the first V, means that big data needs a certain size to be big. Such high volume mainly stems from data that is produced not just once, but again and again, perhaps as a result of an experiment, such as the Large Hadron Collider, or through repeated re-transmission, as for example in an e-mail conversation. It is this re-transmission which helps determine the second V, or Velocity. The third V describes how data assets of differing format or origin can nevertheless be exploited as big data when combined together in information repositories. For example, the family history service Ancestry had over fourteen billion records in August 2014 and was adding two million daily.[10] However, these records are taken from many different sources and thus show great Variety. While the three Vs of Volume, Velocity and Variety provide a clear conceptual framework, they do not cover every aspect of big data and it has been proposed that a number of other Vs should be added to the mix such as Veracity and Value. Many business and government executives are uncertain how accurate their data is and the level of Veracity can have a big effect on predictions made from data.[11] Where large quantities of accurate data can be used to predict sales movements or other business cycles the data has a monetary Value.[12]

Big data developments frequently attract controversy, often as much for social and cultural reasons as for any practical computing issues. Some extreme claims by proponents of big data have prompted extensive debate. In particular, Chris Anderson in 2008 *Wired* magazine declared the 'end of theory', as he argued: 'This is a world where massive amounts of data and applied mathematics replace every other tool that might be brought to bear. Out with every theory of human behavior, from linguistics to sociology . . . With enough data, the numbers speak for themselves' (Anderson 2008: 1).

The claim of the 'end of theory' was meant as a provocation and may superficially be read as a plea to use the availability of large quantities of data to return to positivistic methodologies. Anderson apparently suggests that research questions should derive not from the broader intellectual concerns of a particular discipline but rather from numerical patterns and anomalies in large data sets. Many commentators see such data-driven research, which looks for patterns without explaining them, as the chief characteristic of big data methods (Mayer-Schönberger and Cukier 2013). A more cautious reading of Anderson's piece might suggest that he was only proposing the abandonment of some theories of human behaviour, particularly sociological ones, and their replacement with mathematically based models. Nevertheless, the idea that focusing on data means that critical and theoretical frameworks can be abandoned has been influential. An article on the digital humanities in the *New York Times* in 2010 echoes Anderson's impatience with critical theory:

> A history of the humanities in the 20th century could be chronicled
> in 'isms' – formalism, Freudianism, structuralism, postcolonialism –

grand intellectual cathedrals from which assorted interpretations of literature, politics and culture spread. The next big idea in language, history and the arts? Data.

Members of a new generation of digitally savvy humanists argue it is time to stop looking for inspiration in the next political or philosophical 'ism' aand start exploring how technology is changing our understanding of the liberal arts. This latest frontier is about method, they say, using powerful technologies and vast stores of digitised materials that previous humanities scholars did not have.[13]

Callebaut (2012) has dismissed as fundamentally flawed the idea that scientific and other research can be data-driven, pointing out that Anderson misrepresents the role of modelling in biological research and emphasising that research questions, methods and interpretation remain fundamentally driven by the current theoretical understanding of the relevant discipline. Callebaut quotes Charles Darwin who declared that 'all observation must be for or against some view if it is to be of any service' (Callebut 2012: 74). The analytical and predictive techniques used in big data are themselves founded on statistical and mathematical theories (Mayer-Schönberger and Cukier 2013: 70–2). The idea that 'raw data' represents an objective factual quarry is an illusion (Gitelman 2013). Much of the disruptive power of big data for traditional fields of inquiry stems not from giving up theory but rather from the way in which very large quantities of data challenge existing methodologies and require researchers to explore the use of new theoretical models, perhaps drawn from other subject areas that had previously seemed alien (Boyd and Crawford 2012). West (2013: 1) declares that 'Big data needs a big theory'.

The second major area of concern about big data is the use of data in surveillance or 'dataveillance', which describes 'the systematic monitoring of people or groups, by means of personal data systems in order to regulate or govern their behavior' (Degli Exposit 2014: 1). Following the revelations by Edward Snowden of the enormous scale to which the intelligence agencies of America and its closest allies harvest e-mail, mobile phone and other communications, big data has become linked in the popular imagination to the perfection of state surveillance, although such forms of big data as meteorological data have little to do with surveillance. Before big data 'all surveillance was inherently partial and analogue in nature and produced varying levels of recorded data, ranging from observations that were unrecorded to detailed logs or continuous recordings, sometimes applied to samples' (Kitchin 2014: 88). The East German secret police, the Stasi, famously collected all information it could get on almost everybody in the German Democratic Republic until the end of that state in 1989. Nowadays, smartphones collect more information about their owners than the Stasi could have dreamt of, and Snowden has revealed how modern security agencies make use of that. Compared to the Stasi records, this data is real-time and delivers valuable information such as location and movement in a format computers can easily use. The Stasi was limited in how deeply it could analyse the data or represent it in graphs. With big data, it has become much easier to analyse relationships and networks in great depth. Moreover, big data can be used for pre-emptive actions by police and security agencies, as Lyon (2014) explains:

> Big Data reverses prior policing or intelligence activities that would conventionally have targeted suspects or persons of interest and then sought data about them. Now bulk data are obtained and data are aggregated from different sources *before* determining the full range of their actual and potential uses and mobilising algorithms and analytics not only to understand a past sequence of events but also to predict and intervene *before* behaviors, events, and processes are set in train. (Lyon 2014: 4)

Another area of controversy around big data is the use of quantification. The role of statistics is regarded with particular suspicion in the humanities, where the use of mathematical modelling and computing in the 1950s and 1960s in such controversial disciplinary developments as 'cliometrics' in history or the 'new archaeology' had caused bitter debates (Fogel and Davis 1966; Clarke 1973). Consequently, big data causes long-standing tensions about the relationship of quantification to more hermeneutical work in the humanities to resurface. Moreover, the use of statistical techniques to predict human behaviour and cultural trends seems antagonistic to many of the values traditionally associated with research in the humanities. Mayer-Schönberger and Cukier declare that 'Predictions based on correlations lie at the heart of big data' (2013: 55). They describe a widely reported case in which the American retailer Target identified two dozen products that enabled the company to calculate a 'pregnancy predictor' score for each customer, which even calculated the due date of the pregnancy. Mailings sent by the store using this predictor revealed the pregnancy of a high-school girl before her parents were aware of it (ibid.: 57–8). Predictive data analytics are now being used for applications ranging from dementia research[14] to improving the content of film scripts.[15]

For many humanities scholars, such techniques are at best irrelevant and at worst dangerous. Adam Kirsch declares that: 'In humanistic study, quantification hits its limits (even if quantifiers refuse to recognise them). It is much easier to measure the means – books published, citations accumulated – than the ends' (Kirsch 2014: 1). For the sociologist Emma Uprichard, big data reuse of methods from physical, engineering, computational and mathematical sciences leads to 'reductionist approaches', is 'deeply positivist' and will finally end up in a 'methodological genocide'.[16] For Uprichard, the enthusiasm for statistical analysis of data offers few intellectual insights or frameworks for addressing grand challenges:

> Let's face it, big data is not going to solve our big social problems, such as global warming, violence, genocide, war, social divisions, sexism, racism, disability, homophobia, water and food security, homelessness, global poverty, health and educational inequality, infant mortality, care for the elderly, and so on.[17]

This chapter will attempt to counter such anxieties about the role of big data in the humanities by focusing on approaches which, by being firmly grounded in the traditional values of humanities disciplines, enhance existing methods to produce fruitful humanities research. Big data poses many methodological challenges, but these pressures should prompt humanities scholars to pay much closer attention to methodological issues than they have in the past.

## DATAFICATION OF THE HUMANITIES

Many very large new data sets have been created in the past decade for humanities scholars. For example, as of the end of 2013, the European Union Cultural Heritage aggregator Europeana had made available over thirty million digital objects through its portal.[18] If big data is simply the extension of data from the giga- and terabyte domains into the peta- and exabyte (and beyond), then the humanities already deals with such big data. While nothing matches the 200 petabytes generated by the Large Hadron Collider in the search for the Higgs Boson, 'big humanities' can nevertheless rival scientific data in size (Hand 2011). The Sloan Digital Sky Survey, for instance, had brought together about 100 terabytes of astronomical observations by the end of 2010. This is big data, but not as big as some humanities data sets. The Holocaust Survivor Testimonials Collections by the Shoa Foundation contained at the same time 200 terabytes of data. The American English data set from Google Books contain 200 billion words,[19] while another typical digitisation project, the Taiwanese TELDAP archive of Chinese and Taiwanese prints, images and other heritage objects, had over 250 terabytes of digitised content in 2011.[20] This is not surprising considering that most of these collections have multimedia files which tend to be bigger than other types of data. Where videos and images dominate, tera- and petabyte sets are readily produced.

This kind of multimedia data can be entertaining for humans, but computers have problems processing it because it is unstructured. Without going into too much detail, in computing, unstructured data refers to text, images, etc., while a good example of structured information is a spreadsheet. A good analogy for understanding the difference between structured and unstructured data is comparing the way in which an address book organises information (structured) with a box full of unlabelled photographs (unstructured). Structured data is often organised in relational databases which follow a standard that has not significantly changed over the last twenty years. It is called Structured Query Language (SQL) and is based on analysing and processing sets of data; its return produces sets of records that comply with certain conditions. So, for instance, a typical query would return all records of people born on 28 August 1975. Organising data so it can be interrogated by SQL can be very laborious, and the conventions used to label the information in SQL are often difficult for humans to work with. However, for computers this kind of data is easier to understand because there are unambiguous identifiers describing the information in the record or how records can be joined up.

If the data is unstructured or unstructured queries are allowed, more computational intelligence needs to be added to the application so that the unstructured information can be transformed into something a computer can process. A commonly used method to transform unstructured into structured information is information extraction (Cowie and Lehnert 1996). Google mail, for instance, identifies what it believes to be dates and times for appointments in e-mails and lets the user add those directly to his or her calendar. Extracting information such as appointment details in e-mails is never a perfect process, but it helps with the organisation of e-mail content. Information extraction (IE) does not target the whole document (like the content of an e-mail) but individual facts found there.[21] IE's aim is therefore to find and process smaller parts of documents and extract facts from them, such as who did what and when. Relationships can also be extracted, for instance that Rome is the capital of Italy. In this way structured data is derived from unstructured information – texts in this case.

There are many openly accessible IE tools now on the web, of which OpenCalais from Reuters is one of the better known. You can use this tool yourself. Simply go to <http://viewer.opencalais.com/> and submit any English text. Figure 11.1 shows the results using the first paragraphs of the Wikipedia entry on Indonesia.

You can see how the OpenCalais tool annotates words in the text such as places or organisations. While this tool is easy to use and potentially has a lot of applications in humanities research, it is immediately obvious from Figure 11.1 that this is not a perfect process and contains mistakes. Current IE systems are very good at identifying documents of a specific type and then extracting information according to pre-defined templates, such as a list of all places. But IE can easily fail if the spelling is slightly different or other factors hide the original name, as in Figure 11.1, where the adjective form of Indonesia is not recognised. While IE is helpful in retrieving information from a particular group of sources, such as news wires, its more extensive deployment in historical texts requires further linguistic research in order to establish how words relate to one another so that computers can be programmed to recognise the different functions of words. Despite IE's limitations, it gives a good idea of how it is necessary to identify information such as names and places if we are to transform the unstructured information in (say) the text of a Victorian novel into data.

Figure 11.1   OpenCalais extraction of information from Wikipedia entry on Indonesia
(Source: The author)

We tend to use the term 'digitisation' very generically. There is a difference between a book which has been scanned to produce a digital image which has been made available as unstructured data, and a text which has been keyboarded and marked up for automated processing so that it is structured data. The process of creating structured data has been described as 'datafication' (Mayer-Schönberger and Cukier 2013). Mayer-Schönberger and Cukier (2013) define data as everything that can be digitally repurposed and analysed by machines. For Mayer-Schönberger and Cukier, 'to datafy a phenomenon is to put it in a quantified format so that it can be tabulated and analysed' (Mayer-Schönberger and Cukier 2013: 78). This is different from the process of producing a digital surrogate based on digitising originally analogue content by (for example) transferring a microfilm of a book to digital form or making an MP3 version of a taped interview. Mayer-Schönberger and Cukier (2013) rightly point out that big data is not related to the kind of digitisation work which libraries and museums undertake to provide access to their collections, but rather to the desire to produce quantifiable pieces of data a computer can ask relevant questions against.

The concept of 'datafication' has particular significance in the humanities, since it suggests that for big data work unstructured data is not enough. This would mean that the digitised files of sound recordings of interviews of Holocaust survivors in the Shoa Foundation described above are not big data as such. If the large collections of images, sound or video commonly used by humanities researchers are not big data, then what is? For Mayer-Schönberger and Cukier (2013), the Google Ngram Viewer is a perfect example of how humanities content can be 'datafied' by splitting it up into n-grams or smaller chunks of data. Ngrams are here simply 'n' letters in a word joined together. The word 'data', for instance, contains two 3-grams: 'dat' and 'ata'. N-grams are often used in linguistic analyses to counteract the challenges of heterogeneous data, if, for instance, texts generated by automated optical character recognition packages contain recognition errors. N-grams help with processing those words with inaccuracies. N-grams might not help humans understand texts better but they definitely provide computers with a way to parse large amounts of inconsistent and incoherent texts. Other examples of large sets of structured data used in the humanities are library catalogue records or linguistic corpora. Like n-grams, these data sets are sufficiently large that they cannot be effectively deployed without sophisticated computational methods.

## METHODS

The computational methods associated with big data, far from undermining the humanities, can revitalise long-standing methods used by humanities scholars and give them renewed relevance. Prior to the computer, scholars laboured with difficulty to compile Domesday statistics and geographies from the vast inventory of information about eleventh-century England preserved in the Domesday Book. The Domesday Book has now been 'datafied', with all the information placed in a database so that it can be searched, analysed and mapped, allowing landholding patterns, values, estate structures and much else to be analysed.[22] This work on the Domesday Book reflects the long-standing interest of humanities scholars in analysing the membership of historical populations and groups. Prosopography is the writing of a collective biography of groups that share certain attributes and is 'a means of profiling any group of recorded persons linked by any common factor' (Magdalino 2003: 43). Prosopography investigates 'what the analysis of the sum of data about many individuals can tell us about the different types of connection between them, and hence about how they operated within and upon the institutions – social, political, legal, economic, intellectual – of their time' (Keats-Rohan 2000: 2). According to the Byzantine scholar Paul Magdalino prosopography is 'a powerful analytical tool which literally reduces history to atoms, for a *prosōpon* is an *atomon*, the indivisible unit of human existence' (Magdalino 2003: 46).

The compilation of prosopographies has a venerable history in the humanities, but their publication in big printed volumes makes them both cumbersome to use and expensive. The *Prosopography of the Later Roman Empire* (Jones et al. 1980), which contains biographies of much of the recorded population of the late Roman world from CE 260 to 641, was published in three volumes containing altogether 4,157 pages and costing over $1,000. A reviewer commented: 'The book is very fat, and the sheer quantity of matter it contains is an impediment to judicious assessment' (Bowersock 1976: 85). The 'datafication' of a volume like this not only makes it easier to use but also allows the information it contains about individuals to be explored in new configurations so that networks, patronage and patterns of movement can be reconstructed. A project has recently been started to digitise the *Prosopography of the Later Roman Empire*.[23]

Prosopographical studies are used for many different cultures and societies. A project at the University of California Berkeley is developing a prosopography from a set of Hellenistic Babylonian legal texts in ancient cuneiform script, and is using this to test an open-source prosopographical toolkit that generates interactive visualisations of the biological and social connections that individuals documented in legal and administrative archives.[24] The Department of Digital Humanities at King's College London has developed a series of prosopographical projects which cover the populations of the Byzantine Empire, Anglo-Saxon England and medieval Scotland as well as more modern groups such as the Clergy of the Church of England (Bradley and Short 2005; Bradley and Pasin 2013). The databases developed by those projects have been used to examine such major research questions as the development of the sense of national identity of the lowland Scottish population and the operation of social networks in medieval society (Hammond 2013: 3–4). The potential for prosopography to contribute to the humanities work in Big Data was recognised by the Arts and Humanities Research Council in the UK when it funded as part of its portfolio of Big Data projects two prosopographical projects in such different domains as Greco-Roman histories and archival research.[25]

Because prosopographies catalogue all recorded persons from a particular country or time period, they provide insights into the ordinary people often overlooked in historical research. Magdalino describes prosopographies as being 'like a police file' (2003: 47) and the same analytic techniques that cause concern when used by security agencies or by retailers like Target can be used to identify trends, networks and social structures in historic populations. One of the best-known digital humanities projects is the Proceedings of the Old

Bailey, which created a searchable edition of over 197,000 trials at the central criminal court in London between 1674 and 1772, containing over 127 million words of text.[26] The Old Bailey Proceedings contain biographical details of approximately 2,500 men and women executed at Tyburn and is the largest body of texts concerning the lives of non-elite people ever published. The Old Bailey data has been used to analyse the life and behaviour of beggars in eighteenth-century London (Hitchcock 2005). A data-warehousing tool is provided which enables graphing and visualisation of data from the trials.[27] In a project called 'Data Mining with Criminal Intent', sophisticated text mining and statistical software were used to investigate the word count for each individual trial. Changes in the length of reports suggests that major changes in the conduct and function of the trial occurred in the nineteenth century, with the increasing use of counsels and the rise of adversarial proceedings (Cohen et al. 2011). An example of the potential synergies between prosopographical studies and big data analytics is a study which uses the Jensen–Shannon divergence, a popular method of measuring the similarity between two probability distributions, to show how a distinction emerged between violent and non-violent crimes in trials at the Old Bailey in the early nineteenth century (Klingenstein et al. 2014).

The European Holocaust Research Infrastructure (EHRI)[28] investigates, collects and integrates Holocaust material from archives across Europe. Again, the aim is to link the material in archives into virtual collections so that, for example, prisoners of Auschwitz can be traced with regard to their countries of origin, other camps where they were prisoners, etc. This once more illustrates how the linking, identification and cross-referencing of records which characterises big data is also at the heart of much traditional historical research. The EHRI illustrates the potential difficulties of building such links. For the Holocaust, a great deal of work on identifying and also preserving documentation on the Holocaust has already taken place, mostly by Yad Vashem in Israel[29] and the United States Holocaust Memorial Museum (USHMM).[30] Both have collected huge numbers of Holocaust documents. Initially this was done by photocopying, more recently by digitisation. Yad Vashem and USHMM hold massive copy collections in both analogue and digital form, and the EHRI's work in integrating material from these archives has to discriminate carefully not only between copies and originals, but also between older copies and more recent copies. Consequently, the EHRI has to reconstruct how the copy collections were built up. The EHRI shows how, in building up large resources that survey and link material from different collections, it is essential to consider the provenance and status of individual records, disentangling copies and originals.

The EHRI is undertaking experiments with social network analysis, which is used in the social sciences to analyse the interaction between individuals or groups. This generally involves the application of quantitative methods to represent communities as networks, in which the nodes correspond to individuals and the links between nodes to the relationships between individuals (Easley and Kleinberg 2010). (The exciting field of social network analysis is discussed in more detail in Robert Glenn Howard's chapter in this volume.) One of the key features of social network analysis as described by Easley and Kleinberg (2010) is the idea of strong and weak ties between actors in social networks. In the EHRI, weak ties could be indicated by fewer historical relationships (by, for instance, living in different communities or belonging to different temples) between two actors, while strong ties require more exchanges. One might assume that those with the strongest ties are also those who help each other make decisions, but this is not necessarily the case. If one runs out of ideas to solve a certain problem, one needs new perspectives. Under such circumstances, it might be necessary to activate the weak ties to connect to new contexts. Weak ties are often more important than strong ties to keep clusters of actors together (Easley and Kleinberg 2010), so that weak ties might be more important in (say) finding a new job, because they will know about a different set of job opportunities than will strong ties. For the EHRI, we are investigating whether weak ties helped people to survive during the Holocaust. In which network did Jews take part that supported their survival? How did they participate in these networks? How were these networks structured nationally and internationally?

EHRI reminds us that not everything is digital yet. The coordinator of the project is the Dutch War and Genocide research institution NIOD in Amsterdam. Their archives are 2.5 kilometres long but only 2 per cent of them are available in digital format and then not always accessible online. The objective for 2016 is for 7 per cent of the collections to be available in digital format. NIOD is a well-funded archive and is not alone in that the vast majority of its collections remain undigitised. Recent figures for the British Library, for instance, reveal that only 5 per cent of their collections are digitised. Any kind of big data project in the humanities that aims to link across archives therefore needs to find ways of uniting analogue and digital information. To this end, the EHRI has a twofold strategy of providing virtual access to Holocaust archives through its portal[31] as well as physical transnational access to archival material through fellowships.

If we want to undertake humanities research using only digital material then we need to restrict our research to so-called 'born-digital' material, that is data such as websites or tweets that were created digitally and have never had an analogue form. As already mentioned, such born-digital material is attractive for data analytics work because it is not necessary to undertake the expensive and time-consuming process of turning the information into digital form first. Tweets offer many opportunities for historians or literary scholars to explore social and cultural trends (Kirschenbaum 2012) while web archives will form a major primary source for future research in the humanities (Brügger and Finnemann 2013). One born-digital medium whose potential is sometimes underestimated but which will offer great opportunities for historical and literary research in the future are e-mail archives. The George W. Bush Presidential Center, for example, holds in its collections 200 million White House e-mails of the second Bush administration from the Executive Office of the President.[32] The Bush e-mails are subject to the provisions of the US Presidential Records Act and are only just becoming available for access under the US Freedom of Information Act so we do not know what kind of insights they will produce. In analysing such a vast amount of material, we will probably want to look at patterns of communications and networks of activity, in just the same way that Edward Snowden has shown that the National Security Agency analyses metadata from the e-mails of the general population: '[H]istorians can look for clusters of emails around various events and see, perhaps, the discussions that went on and the thinking and the mindset of individuals in the White House during the various stages of those big events.'[33] We could use the number of emails an individual sends out to estimate his or her influence. It is probably then also a good idea to investigate the patterns of replies to those e-mails. The from–to relationship is part of all e-mail communication metadata, as is the time an e-mail was sent. We can thus easily identify at which time of day most e-mail communications are happening, which might, for instance, indicate how work on particular policies or legislation proceeded. We could even perform 'sentiment analysis' on words used in e-mails and produce statistical analyses of the mood in the White House at different times.

This kind of research has been done with the publicly available Enron data set, which contains e-mails made public during the legal investigation into the collapse of the Enron corporation (Klimt and Yang 2004). Russell (2013) gives detailed instructions on how this e-mail data can be interrogated. Simply counting numbers of e-mails and correspondents and tabulating the frequency of contact can 'tell you so much with so little effort' (Russell 2013). Such simple tabulation, making use of the rich metadata and structured information associated with e-mails, is a good first step in analysing such data sets. It is not difficult, for instance, to track the number of documents sent out at a certain period of time of day during a week. From these groupings it is easy to cluster groups of e-mailers who are involved in frequent, direct communication with each other. Most users would probably not be adventurous enough to download the Enron e-mail archive, but a sense of what is possible can be easily achieved by analysing your own e-mail. MIT have produced a tool called 'Immersion' which analyses the metadata of messages in any g-mail box (precisely the method used by the security agencies) and produces visualisations showing the relationship between correspondents, enabling any user to explore the potential of the data produced by e-mail activity.[34] You can easily and safely try Immersion for yourself.

There are many other aspects to the exploration of e-mail archives which could be discussed here, such as the way in which natural language processing techniques can be used to investigate word occurrences. Russell (2013: 261–4) shows how simple text mining techniques involving the identification of terms such as 'raptor' can be used on the Enron corpus to identify which executives were aware of the use of fraudulent financial instruments. However, in conclusion, we would like to draw attention to one aspect of the study of e-mails which has given renewed relevance to a venerable humanistic method. Diplomatic is the study of the layout and formulae of ancient charters and is essential to establishing their authenticity. Its origins lie in the techniques used in the fifteenth century by the humanist scholar Lorenzo Valla to establish that the Donation of Constantine, a document supporting claims of papal supremacy, was a forgery. The critical tools of diplomatic were further developed by the French monk Jean Mabillon in writing saints' lives in the seventeenth century, and Mabillon wrote a treatise on diplomatic method, *De Re Diplomatica* (1681). The archival theorist Luciana Duranti has pointed out that e-mail presents many similarities to medieval documents because it can be difficult to establish the authenticity of an individual message. In her book *Diplomatics: New Uses for an Old Science*, Duranti argues that the methods of diplomatic provide tools for assessing the authenticity of such born-digital material (Duranti 1998). The techniques developed by the monk Mabillon thus have renewed value in the world of big data.

## CONCLUSION

Some aspects of big data seem at first sight antagonistic to fundamental values of humanities research: the assumption that, if the data set is large enough, inaccuracies and individual peculiarities will not affect the result; the idea that answers will emerge from just looking at the data; an impatience with contextual discussion; and, above all, the claim that behaviour can be mathematically predicted if you have enough data. But, on closer examination, big data methodologies have more in common with traditional research preoccupations of the humanities than might at first be thought. The police file of the prosopographer is not dissimilar to the mobile phone data gathered by the security agencies. Historians wishing to link references to individuals in different archives have concerns similar to security officers trying to trace a suspect across various social media. The seventeenth-century monk Jean Mabillon can help in exploring e-mail archives.

In a blog post called 'Big Data, Small Data and Meaning', Tim Hitchcock, a leading light in the Old Bailey Proceedings project, has expressed concerns about the way in which some historians, in response to big data, advocate a 'macroscope' view of history emphasising the big picture.[35] For Hitchcock, such approaches share with big data the assumption 'that the "signal" will come through, despite the noise created by outliers and weirdness. In other words, "Big Data" supposedly lets you get away with dirty data. In contrast, humanists do read the data; and do so with a sharp eye for its individual rhythms and peculiarities – its weirdness.' Hitchcock is not unsympathetic to big data approaches. Indeed, he collaborated in the use of Old Bailey data to develop a statistical 'bootstrap model' of the sort used in predictive analytics (DeDeo, Hawkins, Klingenstein and Hitchcock 2013). Hitchcock found that 'working with "Big Data" at scale and sharing methodologies with other disciplines is both hugely productive, and hugely fun'. But what Hitchcock missed in these collaborations with mathematicians and scientists was the close reading of a single datum. For Hitchcock, the humanities are as much about the microscope as the macroscope.

In April 2012, the *Guardian* published a map purporting to show British trade routes between 1750 and 1800, under the headline '18th century shipping mapped using 21st century technology'.[36] The map, surprisingly, shows no trade routes connecting with a number of major ports such as Liverpool or Hull. This was because the ships' logs used in the visualisation were from the Royal Navy; no movements of merchant ships are included. The map shows royal naval ship movements, not trade routes – its creator had failed to think about the provenance of his data. The ease with which data can be visualised and transformed can create a suspension of disbelief in which fundamental critical awareness is forgotten. In using large data sets it is essential to be conscious of the origins of each record, as the need to disentangle copies and originals in the EHRI project shows. This point has been eloquently made by Huggett (2014: 3): data are not 'out there', waiting to be discovered; if anything, data are waiting to be created. As Bowker has commented, 'Raw data is both an oxymoron and a bad idea; to the contrary, data should be cooked with care.' Information about the past is situated, contingent and incomplete; data are theory-laden and relationships are constantly changing depending on context.

In reminding us of the importance of the single datum, Hitchcock is warning us not to forget our fundamental critical tools. Hitchcock urges greater awareness and the 'radical contextualisation' of the language of large data sets: 'We need to be able to contextualise every single word in a representation of every word, ever. Every gesture contextualised in the collective record of all gestures; and every brushstroke, in the collective knowledge of all painting.'[37] This is not a quixotic demand since the data sets exist which would allow such links to be made and recent developments in computing are moving in the direction Hitchcock describes. Among the biggest data used in the humanities are corpora used for research into language, such as the *Historical Thesaurus of English* at the University of Glasgow.[38] If big data techniques were used to link thesauri to such resources as the Old Bailey Proceedings, Hitchcock's vision would have been realised and a virtuous circle made between big data and close reading.[39]

## Notes

1. <http://www.theguardian.com/public-leaders-network/2012/apr/18/francis-maude-data-raw-material> (accessed 27 February 2015).
2. <http://blogs.wsj.com/digits/2014/06/06/in-a-single-tweet-as-many-pieces-of-metadata-as-there-are-characters/> (accessed 2 March 2015).
3. <http://twitter.mappinglondon.co.uk> (accessed 1 March 2015).
4. <http://quantifyingmemory.blogspot.co.uk/2013/06/putins-bots-part-one-bit-about-bots.html> (accessed 1 March 2015); <http://ro.ecu.edu.au/isw/54/ (accessed 1 March 2015).
5. <http:// books.google.com/ngrams> (accessed 1 March 2015).
6. <https://digitalcriticaltheory.wordpress.com/2012/03/11/two-different-ways-of-thinking-about-religious-vocabulary/>(accessed 26 February 2015).
7. <http://googlebooks.byu.edu/x.asp> (accessed 6 March 2015).
8. A bit is a binary digit; a byte comprises eight bits; a gigabyte (GB) comprises a million bytes. A terabyte (TB) is 1,000 GB; a petabyte (PB) is 1,000 terabytes; an exabyte (EB) is 1,000 petabytes. A petabyte is enough to store the DNA of the entire population of the United States three times over; an exabyte of storage would contain 50,000 years worth of DVD-quality video. Further up the scale, 1000 exabytes = 1 zettabyte; 1,000 zettabytes = 1 yottabyte; 1,000 yottabytes = 1 brontobyte; 1,000 brontobytes = 1 geopbyte.
9. <http://aalt.law.uh.edu> (accessed 2 March 2015).
10. <http://www.fiercebigdata.com/story/how-ancestrycom-uses-big-data/2014-08-04> (accessed 3 March 2015).
11. <http://www.ibmbigdatahub.com/infographic/four-vs-big-data> (accessed 3 March 2015).
12. <http://www.wired.com/2013/05/the-missing-vs-in-big-data-viability-and-value/> (accessed 3 March 2015).
13. <http://www.nytimes.com/2010/11/17/arts/17digital.html> (accessed 2 March 2015).
14. <http://www.slideshare.net/davidderoure/big-data-for-dementia-research> (accessed 6 March 2015).

35. <http://historyonics.blogspot.co.uk/2014/11/big-data-small-data-and-meaning_9.html> (accessed 7 March 2015).
36. <http://www.theguardian.com/news/datablog/2012/apr/13/shipping-routes-history-map> (accessed 7 March 2015).
37. <http://historyonics.blogspot.co.uk/2014/11/big-data-small-data-and-meaning_9.html> (accessed 7 March 2015).
38. <http://historicalthesaurus.arts.gla.ac.uk> (accessed 7 March 2015).
39. And see now <http://www.gla.ac.uk/schools/critical/research/fundedresearchprojects/samuels/> (accessed 7 March 2015).